\documentclass[preprint,prd,aps,tighten,nofootinbib,amssymb]{revtex4}

\usepackage[dvipdfmx]{graphicx}
\usepackage{epsfig,amssymb,amsmath}
\usepackage{mediabb}
\usepackage{subfigure}
\usepackage[colorlinks=true,linkcolor=blue,citecolor=blue]{hyperref}
\usepackage{float}
\usepackage{here}

\usepackage{ulem}

\makeatletter
\def\Hline{%
\noalign{\ifnum0=`}\fi\hrule \@height 2pt \futurelet
\reserved@a\@xhline}
\makeatother

\def\d{\delta}

\def\h{\theta}

\def\m{\mu}

\def\r{\rho}
\def\s{\sigma}

\def\L{\Lambda}

\newcommand{\beq}{\begin{equation}}
\newcommand{\eeq}{\end{equation}}
\newcommand{\bea}{\begin{eqnarray}}
\newcommand{\eea}{\end{eqnarray}}
\newcommand{\bear}{\begin{array}}
\newcommand {\eear}{\end{array}}
\newcommand{\bef}{\begin{figure}}
\newcommand {\eef}{\end{figure}}
\newcommand{\bec}{\begin{center}}
\newcommand {\eec}{\end{center}}

\newcommand{\la}{\left\langle}
\newcommand{\ra}{\right\rangle}
\newcommand{\ds}{\displaystyle}

\def\lrfp#1#2#3{ \left(\frac{#1}{#2} \right)^{#3}}

\begin{document}
\draft
\tighten
\preprint{TU-1007,~APCTP Pre2015-026,~IPMU15-0181}
\title{\large \bf Level Crossing between QCD Axion and Axion-Like Particle
}
\author{
    Ryuji Daido\,$^{a}$\footnote{email:daido@tuhep.phys.tohoku.ac.jp},
    Naoya Kitajima\,$^{a,b}$\footnote{email:naoya.kitajima@apctp.org},
    Fuminobu  Takahashi\,$^{a,c}$\footnote{email: fumi@tuhep.phys.tohoku.ac.jp}
    }
\affiliation{
$^a$ Department of Physics, Tohoku University, Sendai 980-8578, Japan,\\
$^b$ Asia Pacific Center for Theoretical Physics, Pohang 790-784, Korea,\\
$^c$ Kavli IPMU, TODIAS, University of Tokyo, Kashiwa 277-8583, Japan
}

\vspace{2cm}

\begin{abstract}
We study a level crossing between the QCD axion and an axion-like particle,
focusing on the recently found phenomenon, {\it the axion roulette},  where the axion-like particle 
runs along the potential, passing through many crests and troughs, until it gets trapped in one 
of the potential minima. We perform detailed numerical calculations to determine the parameter 
space where the axion roulette takes place, and as a result domain walls are likely formed. The domain wall network
without cosmic strings is practically stable, and it is nothing but a
cosmological disaster. In a certain case, one can make domain walls unstable and decay quickly 
by introducing an energy bias without spoiling the Peccei-Quinn solution to the strong CP problem.
\end{abstract}

\maketitle

\section{Introduction}
\label{intro}
The QCD axion is a pseudo-Nambu-Goldstrone boson associated with the sponetanous
breakdown of the Peccei-Quinn symmetry~\cite{Peccei:1977hh,Peccei:1977ur,Weinberg:1977ma}.
When the axion potential is generated by the QCD instantons, 
the QCD axion is stabilized at a CP conserving minimum, solving the strong CP
problem.
The QCD axion is generically coupled to photons and the standard model (SM) fermions, 
and its mass and coupling satisfy a certain
relation. On the other hand, there may be more general axion-like particles (ALPs) whose mass and
coupling are not correlated to each other.
The QCD axion and ALPs have attracted much attention over recent decades (see 
Refs.~\cite{Kim:1986ax,Kim:2008hd,Wantz:2009it,Ringwald:2012hr,Kawasaki:2013ae} for reviews), and they are
 searched for at various experiments~\cite{Andriamonje:2007ew,Arik:2013nya,Asztalos:2009yp,Ehret:2010mh,Pugnat:2013dha}.
Furthermore, there appear many such axions in the low-energy effective theory of string 
compactifications,
which offer a strong theoretical motivation for studying the QCD axion and ALPs.

In general, axions have both kinetic and mass mixings\footnote{
There are various cosmological applications of the axion mixing such as
inflation~\cite{Kim:2004rp,Choi:2014rja, Higaki:2014pja, Bachlechner:2014hsa,Ben-Dayan:2014zsa,Higaki:2014mwa} 
and the $3.55$keV X-ray line~\cite{Jaeckel:2014qea,Higaki:2014qua}.}, and 
their masses are not necessarily constant in time.
 In fact, the QCD axion is massless at high temperatures and it gradually acquires a
non-zero mass
at the QCD phase transition. Thus, if there is an ALP with a non-zero mixing with
the QCD axion,
a level crossing as well as the associated resonant transition could occur {\it a la} the MSW effect 
in neutrino physics~\cite{Wolfenstein:1977ue}.  The resonant phenomenon of
the QCD axion and an ALP leads to various interesting phenomena such as suppression of the axion density
and isocurvature perturbations~\cite{Kitajima:2014xla}\footnote{See Ref.~\cite{Hill:1988bu} for an early work on the resonant transition
between axions.}.
Recently, the present authors found a peculiar behavior during the level-crossing
phenomenon: the axion with a lighter mass starts
to run through the valley of the potential, passing through many
crests and troughs, until it is stabilized at one of the potential minima~\cite{Daido:2015bva}. Such axion dynamics is highly sensitive to the initial misalignment angle and
it exhibits chaotic behavior, and so named ``the axion roulette." 
In Ref.~\cite{Daido:2015bva}, however, we studied  the axion dynamics without specifying the
axion masses and couplings, and the application to the QCD axion has not yet been
examined. 

In this paper we further study the level crossing phenomena of axions, focusing on the
mixing between the QCD axion and an ALP. In particular, we will determine the parameter space where the axion
roulette occurs, and domain walls are likely formed. The domain wall network without cosmic strings is practically stable
in a cosmological time scale, and so, it is nothing  but a cosmological disaster~\cite{Preskill:1991kd}. 
We find that, in a certain case,  it is possible to introduce an energy bias to
make domain walls decay sufficiently quickly while not spoiling the Peccei-Quinn
solution to the strong CP problem.

\section{Axion roulette of QCD axion and ALP}
\label{roulette}
\subsection{Mass mixing and level crossing}
\label{levelcrossing}
The QCD axion, $a$, is massless at high temperatures and the axion potential comes from non-perturbative effects
during the QCD phase transition. The QCD axion potential is approximately given by
\beq
V_{\rm QCD}(a) = m_a^2(T)F_a^2\left[1-\cos\left(\frac{a}{F_a}\right)\right]
\label{QCDpotential}
\eeq
with the temperature-dependent axion mass $m_a(T)$~\cite{Wantz:2009it}
\beq
m_a(T)\;\simeq\;
\begin{cases}
\ds{4.05\times10^{-4}~\frac{\L_{\rm QCD}^2}{F_a}\left(\frac{T}{\L_{\rm QCD}}\right)^{-3.34}}& T>0.26\L_{\rm QCD}\\
\ds{3.82\times10^{-2}~\frac{\L_{\rm QCD}^2}{F_a}}& T<0.26\L_{\rm QCD}
\end{cases},
\eeq
where $\L_{\rm QCD}\simeq400~{\rm MeV}$ is the QCD dynamical scale and $F_a$ the decay constant of the QCD axion.
The QCD axion is then stabilized at the CP conserving minimum $a=0$, solving the strong CP problem.

Let us now introduce an ALP, $a_H$, which has a mixing with the QCD axion. Specifically we consider
 the low energy effective Lagrangian,
\beq
\mathcal{L}=\frac{1}{2}\partial_\m a\partial^\m a+\frac{1}{2}\partial_\m a_H\partial^\m a_H-V_H(a,a_H)-V_{\rm QCD}(a)\label{pot}
\eeq 
with
\beq
V_H(a,a_H)=\L_H^4\left[1-\cos\left(n_H\frac{a_H}{F_H}+n_a\frac{a}{F_a}\right)\right],
\eeq
where $F_H$ is the decay constant of the ALP and $n_H$ and $n_a$ are the domain wall numbers of $a_H$ and $a$, 
respectively. We assume that $\Lambda_H$ is constant in time, in contrast to the QCD axion potential.\footnote{
Note that the PQ solution to the strong CP problem is not spoiled by introducing the above potential because those two axions are individually stabilized at the CP conserving minima.}
For later use, we define the effective angles $\theta$ and $\Theta$ by
\begin{align}
\theta &\equiv \frac{a}{F_a},\\
\Theta &\equiv n_H\frac{a_H}{F_H}+n_a\frac{a}{F_a},
\end{align}
which appear in the cosine functions of the axion potential $V_{\rm QCD}$ and $V_H$, respectively.

\begin{figure}[t!]
\centering
\includegraphics[width=8cm]{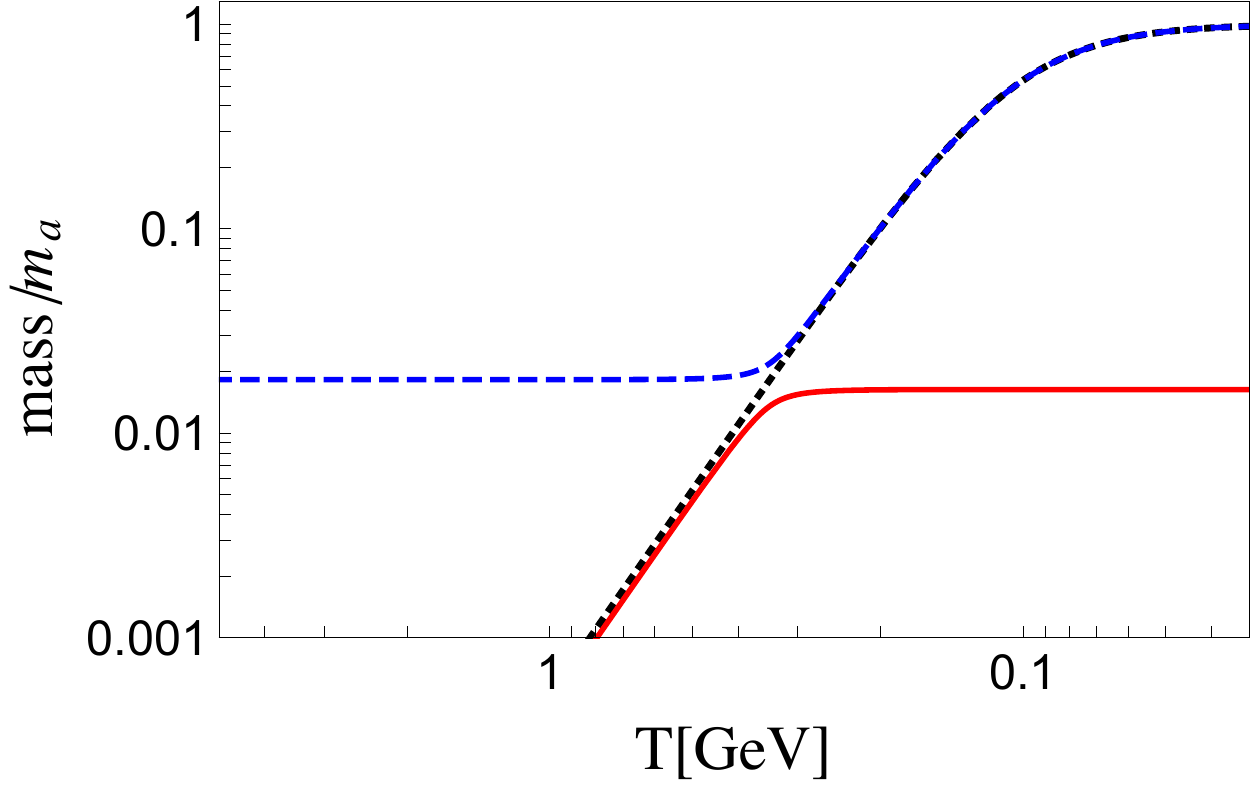}
  \caption{Time evolution of  the two mass eigenvalues, $m_1$ (solid (red)) and $m_2$ (dashed (blue)), and $m_a(T)$ (dotted (black)). We fixed parameters as $f_H=10^{11}$ GeV, $m_H=10^{-7}$ eV, $F_a=10^{12}$ GeV, and $n_a=5$.}
  \label{lc}
    \end{figure}

The mass squared matrix $M^2$ of the two axions $(a_H,a)$ at one of  the potential minima, $a_H=a=0$,  is given by
\beq
M^2=
\Lambda_H^4
\left(
\bear{cc}
\ds{\frac{n_H^2}{F_H^2}} & \ds{\frac{n_H n_a}{F_H F_a}}\\
\ds{ \frac{n_H n_a}{F_H F_a}} &\ds{\frac{n_a^2}{F_a^2}  }\\
\eear
\right)
+
\left(
\bear{cc}
0&0\\
0&m_a^2(T)
\eear
\right).
\label{matrix}
\eeq
Let us denote the eigenvalues of $M^2$ by $m_2^2$ and $m_1^2$ with $m_2 > m_1 \geq 0$.
When $m_a(T) = 0$, one combination of $a_H$ and $a$ is massless, while the orthogonal one
acquires a mass from $V_H$. 
When the QCD axion is almost massless, one can define the effective decay constants $F$ and $f$ for the heavy and light axions, respectively;
\beq
F=\frac{F_H F_a}{\sqrt{n_a^2 F_H^2 + n_H^2 F_a^2}},
\eeq
\beq
f=\frac{\sqrt{n_a^2 F_H^2 + n_H^2 F_a^2}}{n_H}.
\eeq
Even if the heavier axion is stabilized at one of the potential minimum of $V_H$,  the lighter one is generically deviated from
the potential minimum by ${\cal O}(f)$ before it starts to oscillate. 
As the QCD axion mass turns on, the two mass eigenvalues change
with temperature (or time)  (see Fig.~\ref{lc}.).

Now we focus on the case where a level crossing takes place. 
At sufficiently high temperatures, the QCD axion mass $m_a^2(T)$ is much smaller than
any other elements of the mass matrix, and the mass eigenvalues are approximated by
\begin{align}
m_2^2 & \simeq \Lambda_H^4 \left(\frac{n_H^2}{F_H^2}+\frac{n_a^2}{F_a^2} \right) + \frac{\frac{n_a^2}{F_a^2}}{\frac{n_H^2}{F_H^2}+\frac{n_a^2}{F_a^2}} m_a^2(T), \\
m_1^2 & \simeq \frac{\frac{n_H^2}{F_H^2}}{\frac{n_H^2}{F_H^2}+\frac{n_a^2}{F_a^2}} m_a^2(T).
\label{m1b}
\end{align}
In order for the level crossing to take place, the lighter eigenvalue must `catch up' with the heavier one
as $m_a(T)$ increases, i.e., 
\beq
\frac{n_H}{F_H} > \frac{n_a}{F_a}
\label{cond1}
\eeq
must be satisfied. Then, if
\beq
m_a > m_H \equiv \frac{\Lambda_H^2}{f_H},
\label{cond2}
\eeq
is satisfied, 
the level crossing takes place  when the two eigenvalues become comparable to each other,
where we have defined $f_H\equiv F_H/n_H$ and $m_a \equiv m_a(T=0)$.
In this case, the two mass eigenvalues at zero temperature are approximately
given by
\begin{align}
m_2^2 & \simeq m_a^2, \\
m_1^2 & \simeq m_H^2.
\end{align}

In Fig.~\ref{lc}, we show typical time evolution of the two eigenvalues, $m_1$ and $m_2$,
represented by  the solid (red) and  dashed (blue) lines, respectively. The dotted (black) line denotes $m_a(T)$.  
Here we have chosen $f_H=10^{11}$\,GeV, $m_H=10^{-7}$\,eV, $F_a=10^{12}$\,GeV, and $n_a=5$.
At sufficiently high temperatures, one of the combination of $a_H$ and $a$ is almost massless, while the orthogonal combination 
is massive with a mass $\simeq m_H$. As the temperature decreases,  the QCD axion $a$ eventually becomes (almost) the 
heavier eigenstate, while the ALP $a_H$  becomes the lighter eigenstate. 
In terms of the effective angles, $\Theta$ ($\theta$) approximately corresponds to the heavier mass eigenstate 
and $\theta$ ($\Theta$) contains lighter eigenstate well before (after) the level crossing.

The level crossing occurs when the ratio of $m_1$ to $m_2$ is minimized, namely,
\beq
m_a^2(T_{\rm lc}) = \Lambda_H^4 \left(\frac{n_H^2}{F_H^2} - \frac{n_a^2}{F_a^2} \right) \simeq m_H^2,
\eeq
is satisfied, where we have used (\ref{cond1}) in the second equality.
 In the following the subscript `lc' denotes the variable is evaluated at the level crossing. 
During the level crossing, the axion potential changes significantly with time, and the axion dynamics
exhibits a peculiar behavior in a certain case, as we shall see next. 

\subsection{Axion roulette}
There are a couple of interesting phenomena associated with the level crossing between two axions.
First of all, as pointed out in Ref.~\cite{Kitajima:2014xla}, the adiabatic resonant transition could happen
if both axions have started to oscillate much before the level crossing. Then, the QCD axion abundance
can be suppressed by the mass ratio between the two axions. Also if the adiabaticity is weakly broken, 
the axion isocurvature perturbations can be significantly suppressed for a certain initial misalignment angle. 
Secondly, if the commencement of oscillations
is close to the level crossing, the axion potential changes significantly even during one period of oscillation.
As a result, the axion could climb over the potential hill if its initial kinetic energy is sufficiently large. 
The axion passes through many crests and troughs of the potential until it gets trapped in one of the minima,
and we call this phenomenon ``the axion roulette". In the following we briefly summarize the conditions for the axion
roulette to take place.

First, the lighter axion must start to oscillate slightly before or around 
the level crossing. Then the potential changes drastically during the level crossing, and as a result, the axion is kicked into different directions
each time it oscillates. 
\beq
\frac{H_{\rm lc}}{H_{\rm osc}}=\mathcal{O}(0.1-1)\label{condition1},
\eeq
where $H_{\rm osc}$ is the Hubble parameter when the lighter axion starts to oscillate. Before the level crossing, the lighter axion mass 
is approximately given by $m_a(T)$ (see Eqs.~(\ref{m1b}) and (\ref{cond1})), and so,  $H_{\rm osc}$ is basically determined
by  the decay constant $F_a$ and the initial misalignment angle of the QCD axion $a$. 
For $T_{\rm osc}$ and $T_{\rm lc} > 0.26 \Lambda_{\rm QCD}$, the ratio of the Hubble parameters reads
\beq
\frac{H_{\rm lc}}{H_{\rm osc}} = \sqrt{\frac{g_{*}(T_{\rm lc})}{g_{*}(T_{\rm osc})} }
\lrfp{m_H}{m_a(T_{\rm osc})}{- \frac{1}{1.67}},
\label{hratio}
\eeq
where $g_*(T)$ counts the relativistic degrees of freedom in the plasma with temperature $T$.
The condition (\ref{condition1}) can be roughly expressed as 
$ m_H \sim (1-50)\, m_a(T_{\rm osc})$,
and so, the axion roulette takes place for one and half order of magnitude range of the ALP mass.

If the condition (\ref{condition1}) is met, the (lighter) axion gets kicked into different directions around the end points of oscillations
as the potential changes even during one period of oscillation. Therefore, if the initial oscillation energy is larger
than the potential barrier at the onset of oscillations, the axion will climb over the potential barrier.
This is the second condition, and it reads
\beq
\r_{\rm osc}\sim m_{\rm 1}^2f^2>\L_H^4\sim m_{\rm 2}^2F^2 ~~\text{at the onset of oscillations},
\label{condition2}
\eeq
where we have assumed that the initial misalignment angle of the lighter axion is of order unity. 
If the condition (\ref{condition1}) is satisfied, $m_{\rm 1}$ is comparable to $m_{\rm 2}$ at the onset of oscillations. 
Therefore,  we only need mild hierarchy between two effective decay constants, $f > F$.
To this end, one may use the alignment mechanism~\cite{Kim:2004rp}, or one can simply 
assume the mild hierarchy, $F_a > F_H$, which is consistent with (\ref{cond1}) for $n_a \sim n_H$.

As shown in Ref.~\cite{Daido:2015bva}, the axion roulette takes place if the above two conditions are satisfied.
Interestingly, the dynamics of the axion roulette is extremely sensitive to the initial misalignment angle,
and it exhibits highly chaotic behavior (cf. Fig.~\ref{chaotic}). Therefore, domain walls are likely formed 
once the axion roulette takes place. In contrast to the domain wall formation associated with spontaneous breakdown of an approximate
U(1) symmetry, there are no cosmic strings (or cores) in this case. The domain wall network without cosmic strings 
is practically stable in a cosmological time scale, since holes bounded by cosmic strings need to be created on
the domain walls, which is possible only through  (exponentially suppressed) quantum tunneling processes. In the next section we will 
determine the parameter space where the axion roulette takes place.

\section{Numerical calculations of axion roulette}
\label{nume}
Now we numerically study the level-crossing phenomenon between the QCD axion and an ALP.
Specifically we follow the axion dynamics with (\ref{pot}) around the QCD phase transition
 in the radiation dominated Universe.
In order to focus on the dynamics of the lighter axion, 
we choose an initial condition such that, well before the level crossing, 
the lighter axion is deviated from the nearest potential minimum by ${\cal O}(1)$, while
the heavier one is stabilized at one of the potential minima,\footnote{In fact, our main results remain valid even in the presence of coherent oscillations of the heavier axion field~\cite{Daido:2015bva}. }
\begin{align}
\theta_i &= {\cal O}(1),\\
\Theta_i &= 0.
\end{align}
Here and in what follows, the subscript $i$ ($f$) denotes that the variable is evaluated well before  (after) the level crossing. 

The lighter axion $(\theta)$ first starts to move toward the potential minimum, when $m_a(T)$ becomes comparable to the Hubble
parameter. As we assume that this is close to the level crossing (cf. (\ref{condition1})),
the potential changes significantly even during the first oscillation, and the axion is kicked into different directions. As a result,
$\Theta$ (or $a_H$) starts to evolve with time. Note that $\Theta$ corresponds to the lighter axion  after the level crossing.
On the other hand, $\theta$ does not evolve significantly and typically it settles down at the nearest potential minimum 
as it corresponds to the heavier axion after the level crossing.

\begin{figure}[t]
\centering
\begin{tabular}{cc}
\subfigure[$m_H=10^{-8.5}$ eV]{
\includegraphics[width=7cm]{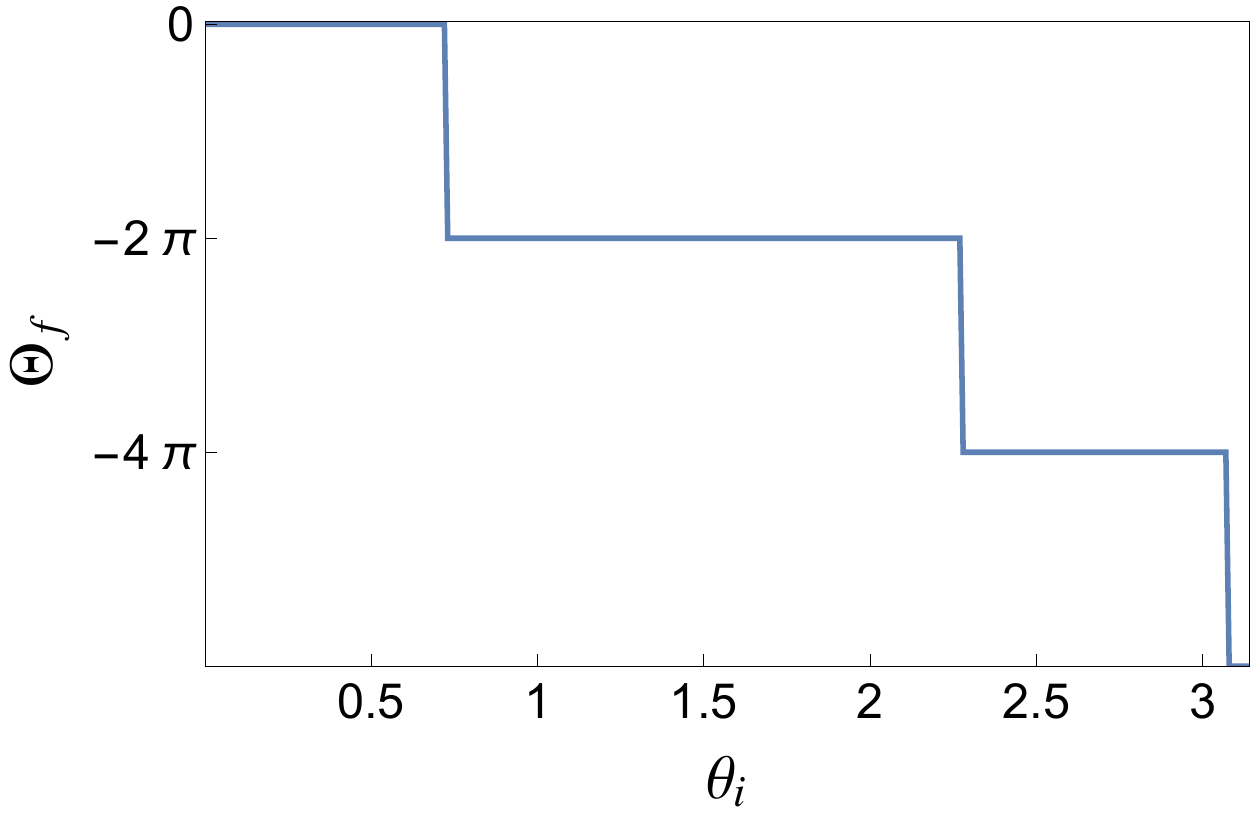}
 \label{sensitivity1}
} &
 \subfigure[$m_H=10^{-7.5}$ eV]{
\includegraphics[width=7cm]{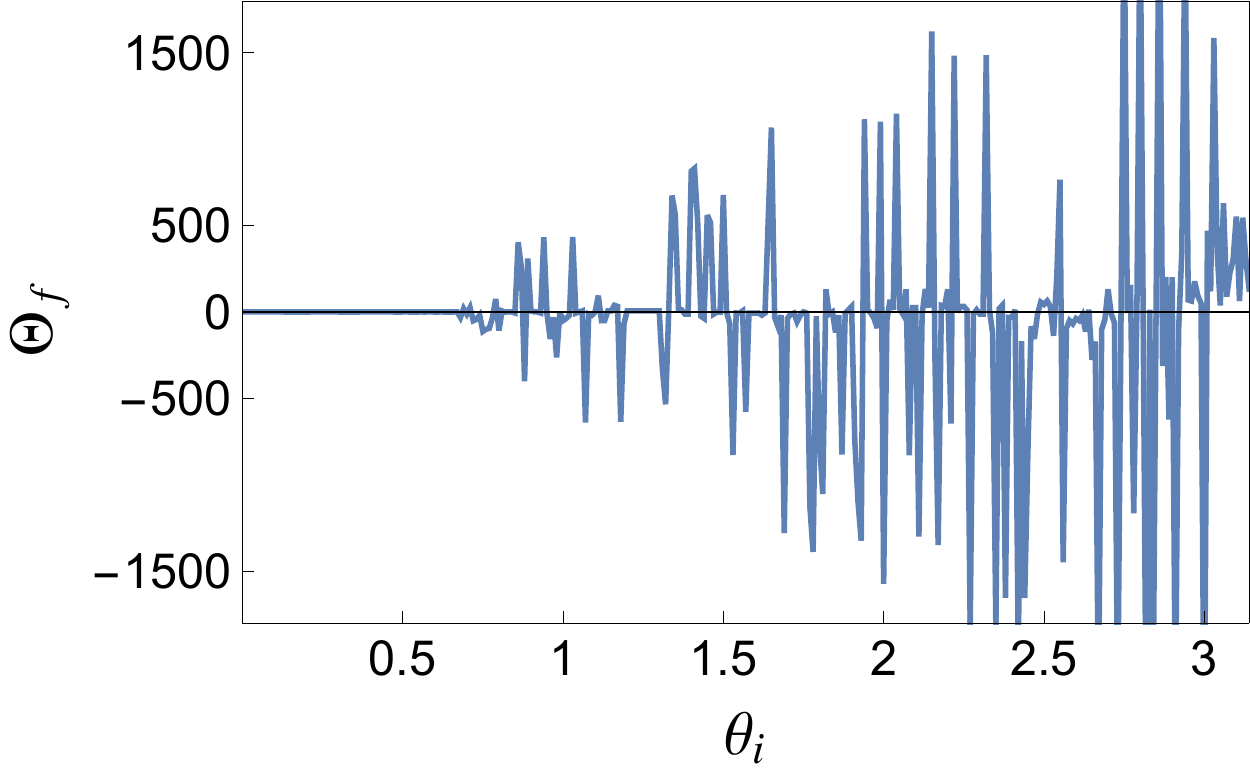}
  \label{sensitivity2}
   } \\
\subfigure[$m_H=10^{-6.5}$ eV]{
\includegraphics[width=7cm]{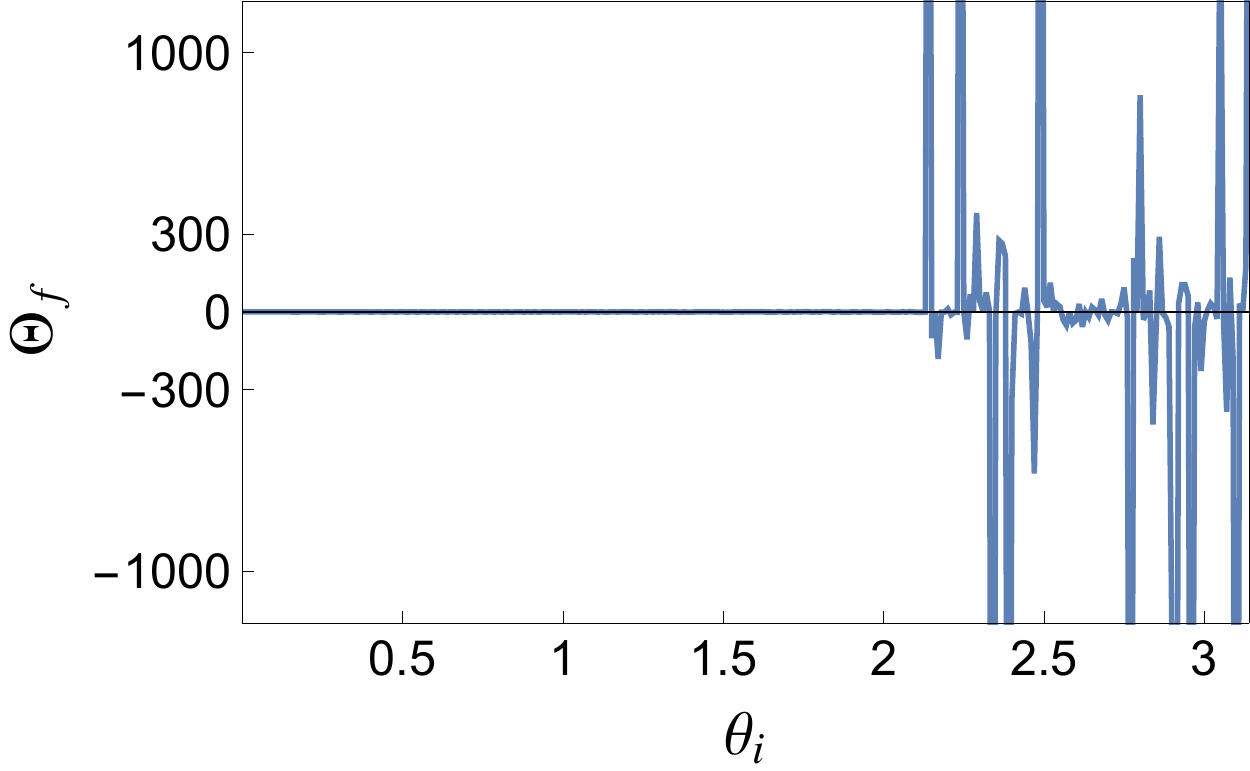}
 \label{sensitivity3}
} & \subfigure[ $m_H=10^{-5.5}$ eV]{
\includegraphics[width=7cm]{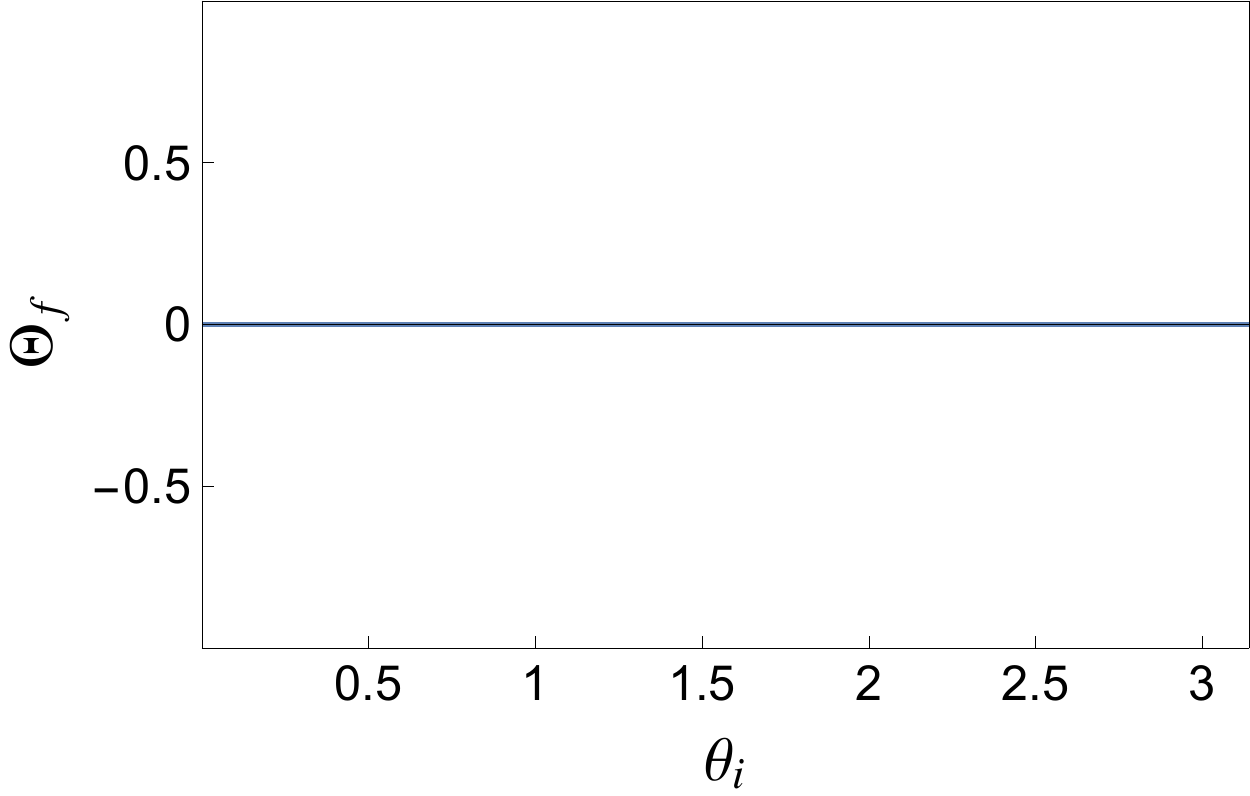}
  \label{sensitivity4}
  }
  \end{tabular}
  \caption{Final values of $\Theta$ are shown as a function of the initial misalignment angle
  for $m_H = 10^{-8.5}$, $10^{-7.5}$, $10^{-6.5}$ and $10^{-5.5}$\,eV. We set $f_H=10^{10}$\,GeV, 
  $F_a=10^{12}$\,GeV and $n_a=5$. }
\label{chaotic}
  \end{figure}

We show in Fig.~\ref{chaotic} the final value of  $\Theta$ for different values of $m_H$ 
as a function of the initial misalignment angle $\theta_i$. Here we have fixed $f_H = 10^{10}$\,GeV,
$F_a = 10^{12}$\,GeV, and $n_a = 5$,  for which $m_a(T_{\rm osc}) \sim 10^{-8}$\,eV and 
$m_a(T=0) \simeq 6 \times 10^{-6}$\,eV.
From Fig.~\ref{chaotic} one can see
that $\Theta_f$ is  extremely sensitive to the initial misalignment angle, and
the axion dynamics exhibits highly chaotic behavior.  
We have confirmed  that $\Theta_f$ takes different values
even if $\theta_i$ differs only by about $10^{-5}$. This sensitivity is considered to arise from the hierarchy
between the initial kinetic energy and the height of the potential barrier.
One can also see that the axion roulette does not occur for the
ALP mass much heavier (or lighter) than $m_a(T_{\rm osc})$.
%
%

\begin{figure}[t!h!]
\centering
\begin{tabular}{cc}
\subfigure[$\h_i=1.5$, $n_a=5$]{
\includegraphics[width=7cm]{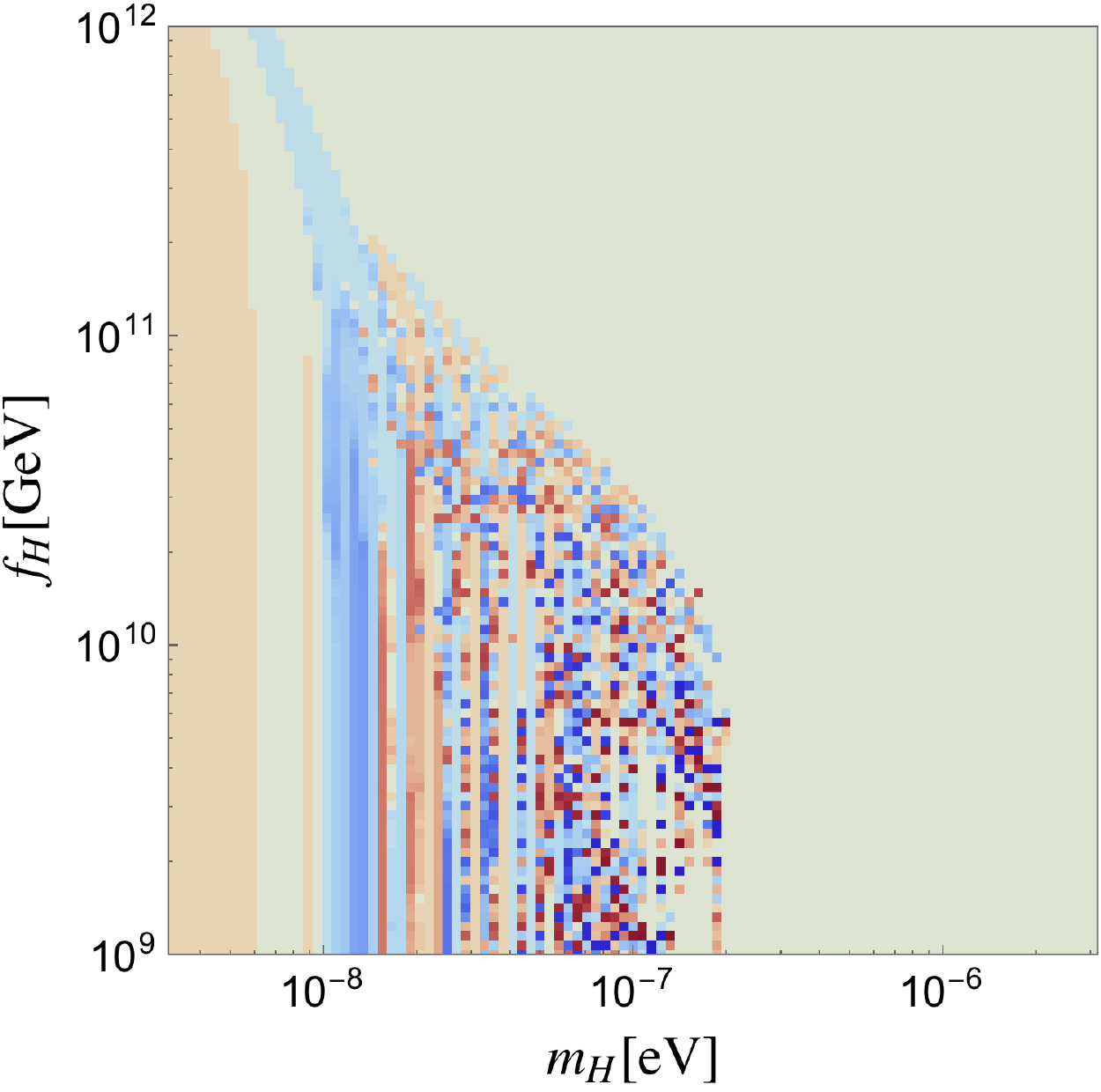}
  \label{fig1}
  } &
\subfigure[$\h_i=2.5$, $n_a=5$]{
\includegraphics[width=7cm]{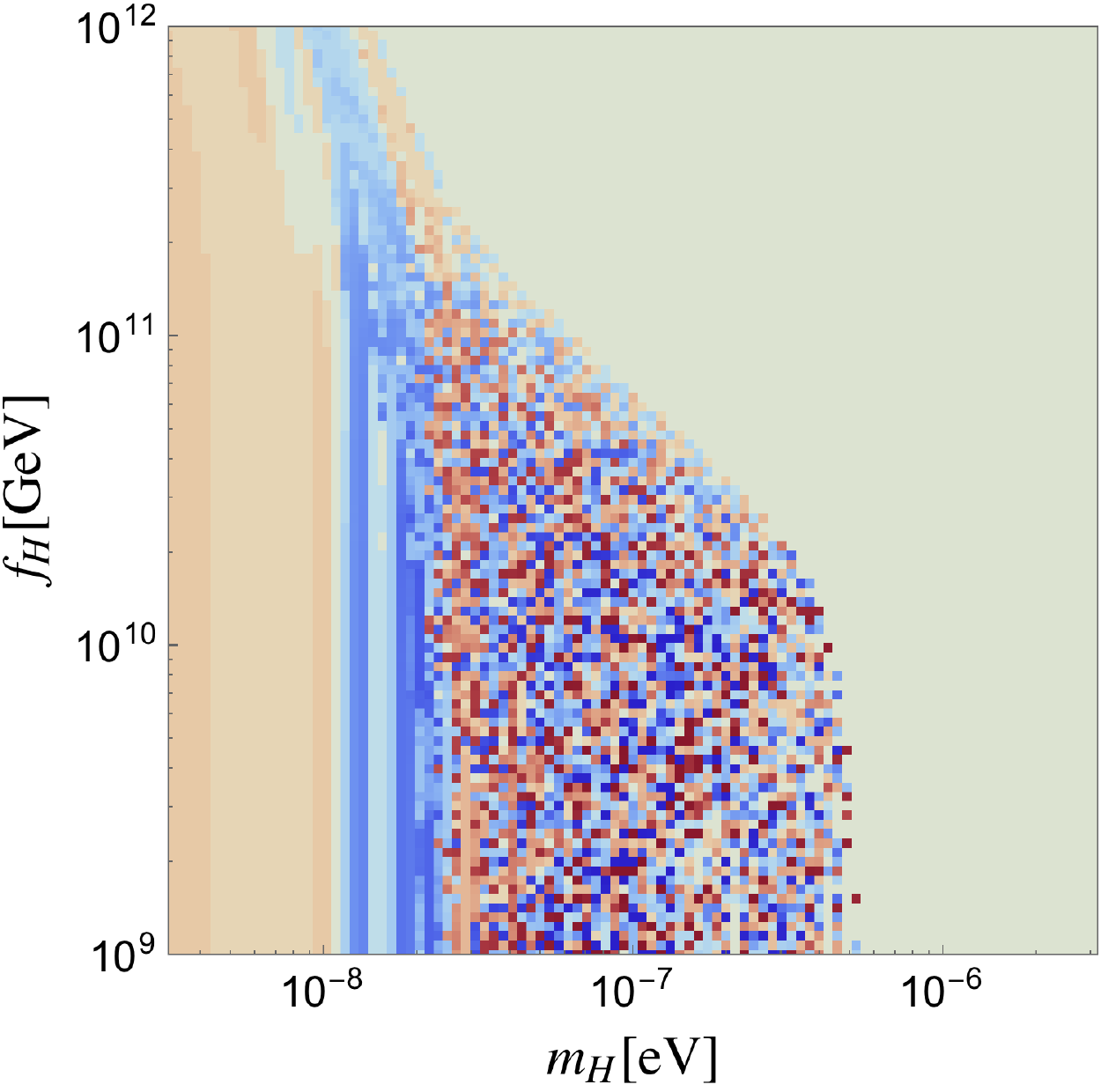}
 \label{fig2}
 } \\
 \subfigure[$\h_i=1.5$, $n_a=15$]{
\includegraphics[width=7cm]{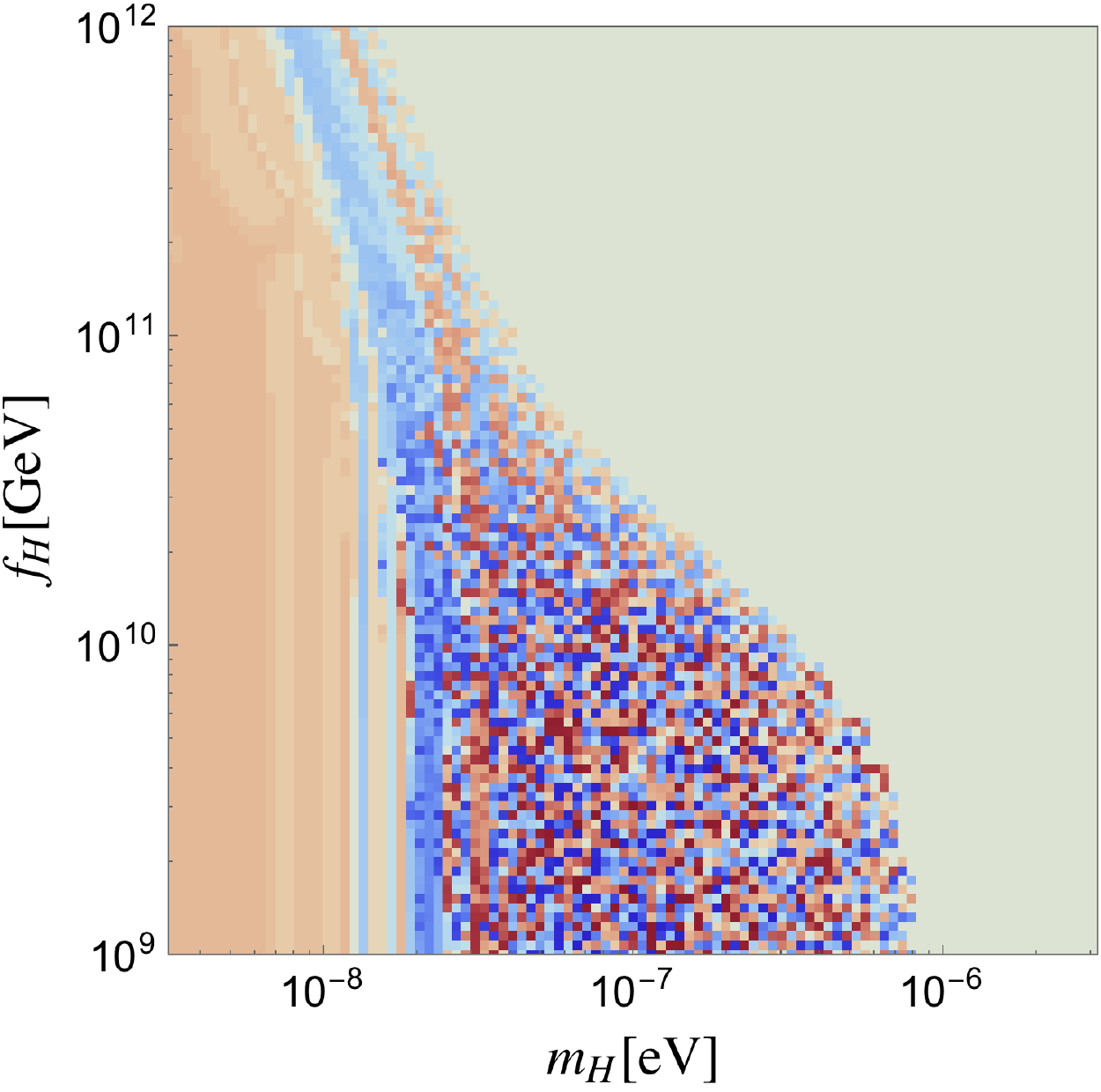}
  \label{fig3}
  }&
   \subfigure{
\includegraphics[height=6cm]{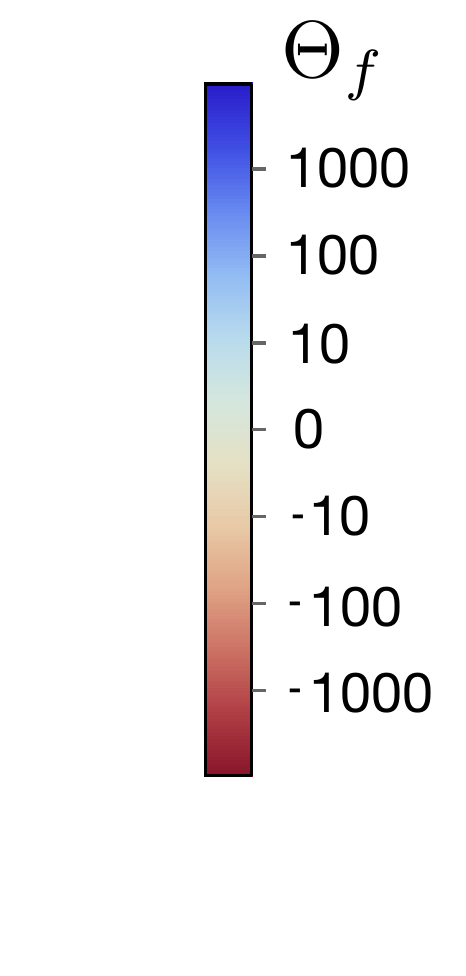}
  \label{density1}
  }  
  \end{tabular}
 \caption{
 The final value of $\Theta_f = a_{H,f}/f_H+n_a a_f/F_a$ are shown by the color bar in the 
 $m_H$-$f_H$ plane. The axion roulette takes place in the multicolored region where
 $\Theta_f$ is highly sensitive to $m_H$ and $f_H$. We set $F_a = 10^{12}$\,GeV and 
 $(\theta_i,n_a) = (1.5,5), \,(2.5,5)$ and $(1.5,15)$, for which $m_a(T_{\rm osc}) \simeq
 7 \times 10^{-9},\,9 \times 10^{-9},\,7 \times 10^{-9}$\,eV, respectively. 
 }
  \label{Fa12}
\end{figure}

In Fig.~\ref{Fa12}, we show the final value of $\Theta$ by the color bar 
in the $(m_H, F_H)$ plane for different values of $\theta_i$ and $n_a$. 
Here we have set $F_a=10^{12}~{\rm GeV}$. The axion roulette takes place
in  multicolored regions where $\Theta_f$ takes large positive or negative values.
In order for the level crossing to take place, $f_H$ is bounded above as $f_H < F_a/n_a$ (see (\ref{cond1})),
which reads $f_H \lesssim 2 \times 10^{11}$\,GeV in Figs.~\ref{fig1} and \ref{fig2},
and $f_H \lesssim 7 \times 10^{10}$\,GeV in Fig.~\ref{fig3}, respectively. These conditions are
consistent with  boundaries of the multicolored regions. Also, the left and right boundaries (the lower and higher 
end of $m_H$) of the multicolored region are determined by (\ref{condition1}). In the right region
the adiabatic transition {\it a la} the MSW effect takes place as long as  (\ref{cond2}) is satisfied. (The condition
 (\ref{cond2}) is outside the plotted region.) 
In the left region, the level crossing takes place before $a$ starts to oscillate, and so, it has no impact on the axion
dynamics. 

Comparing Fig.~\ref{fig1} and Fig.~\ref{fig2}, one notices that
 the multicolored region extends to larger values of $m_H$ as the initial misalignment angle increases 
 from $\h_i=1.5$ to $\h_i=2.5$.
This can be understood as follows. The onset of oscillations is delayed as $\theta_i$ approaches to $\pi$, which increases
the initial oscillation energy, making it easier to climb over the potential barrier. Since the potential barrier is
proportional to $m_H^2$, the axion roulette takes place for larger values of $m_H$. 
Compared to Fig.~\ref{fig1}, the multicolored region in Fig.~\ref{fig3} is extended to larger values of $m_H$.
This is because, as $n_a$ increases,  the effective decay constant $F$ becomes smaller, which makes
the potential barrier smaller.

\begin{figure}[t!h!]
\centering
\begin{tabular}{cc}

\subfigure[$\h_i=1.5$, $n_a=5$]{
\includegraphics[width=7cm]{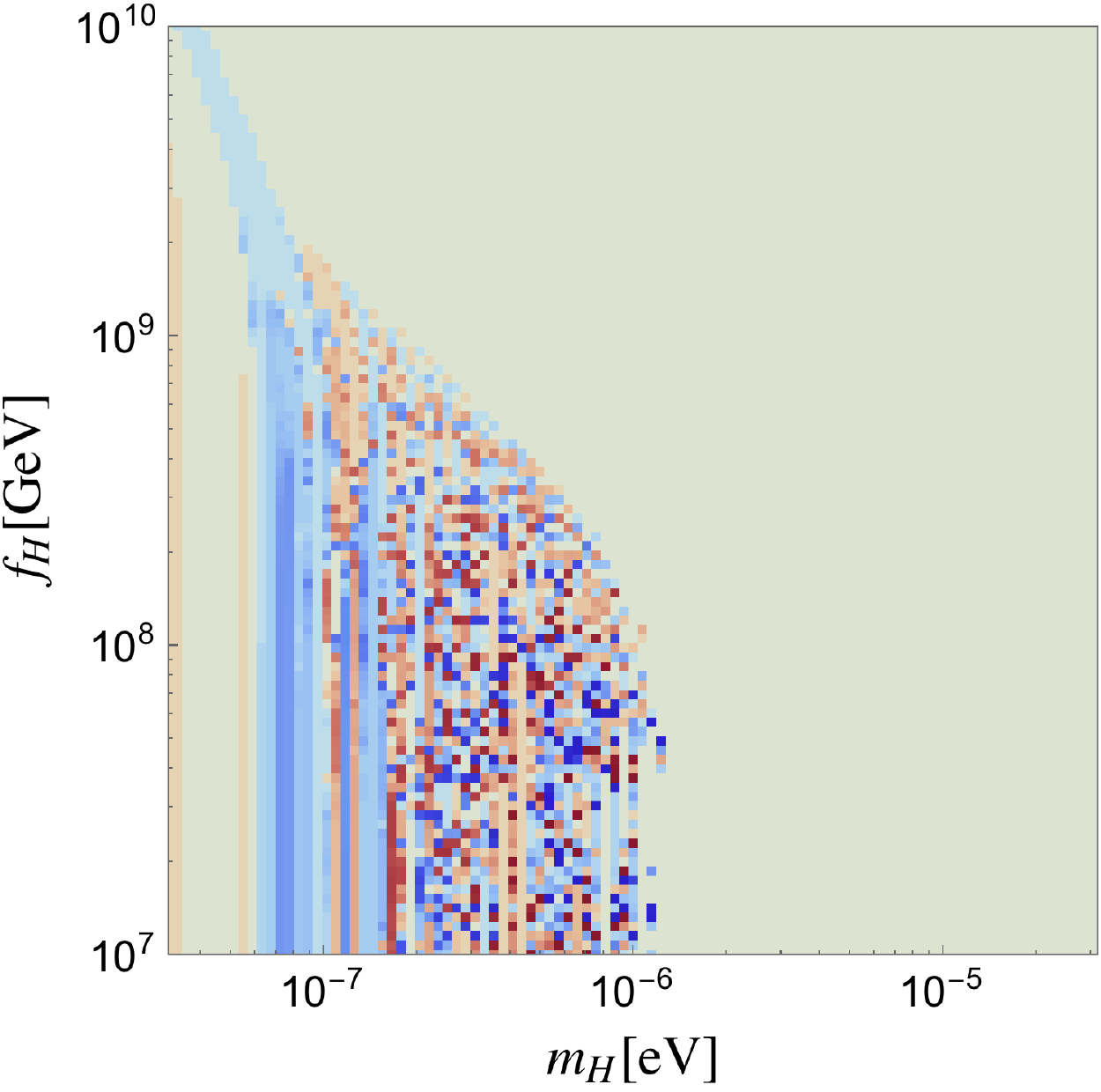}
  \label{fig4}
  } &
\subfigure[ $\h_i=2.5$, $n_a=5$]{
\includegraphics[width=7cm]{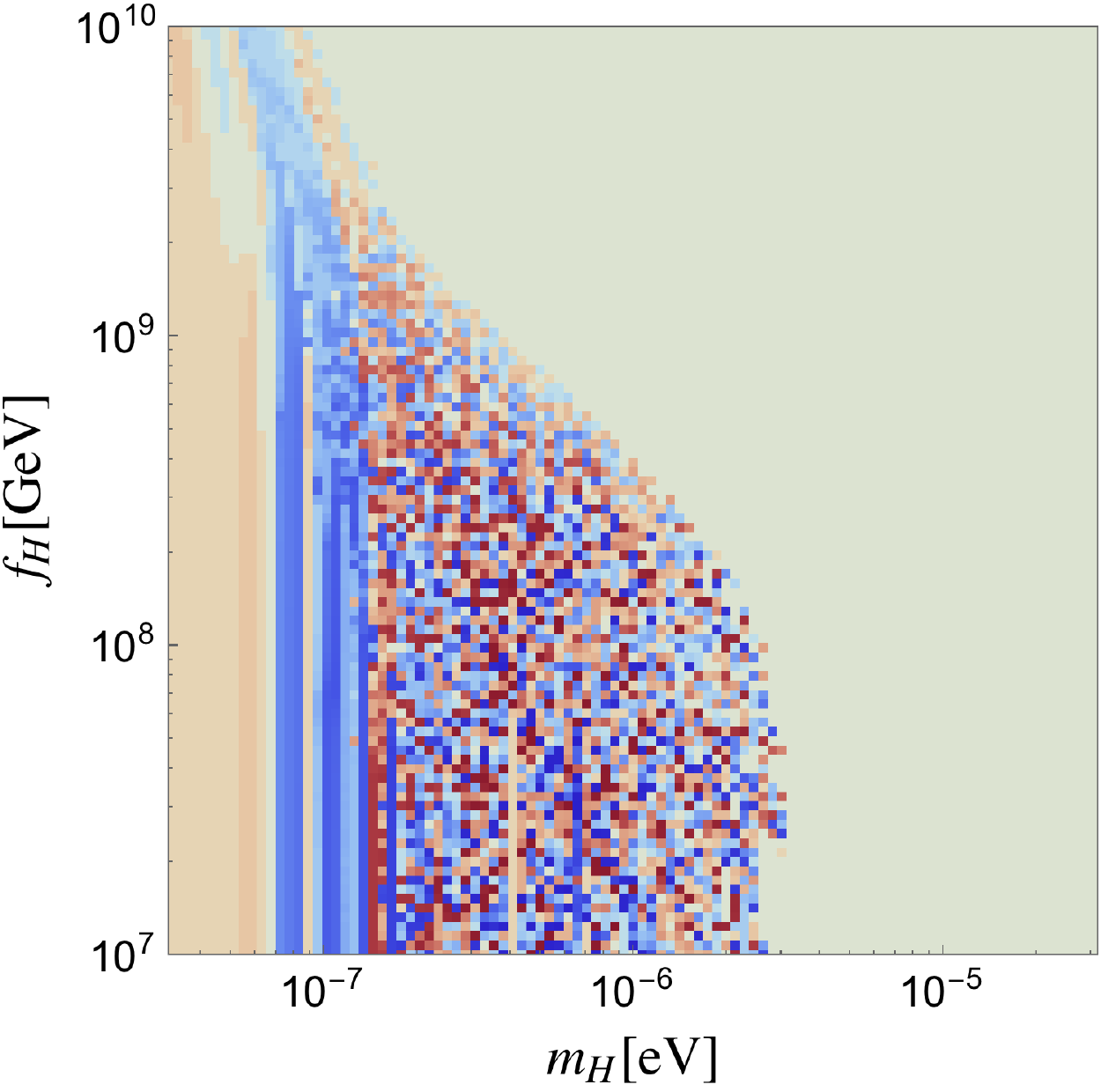}
 \label{fig5}
 } \\
 \subfigure[$\h_i=1.5$, $n_a=15$]{
\includegraphics[width=7cm]{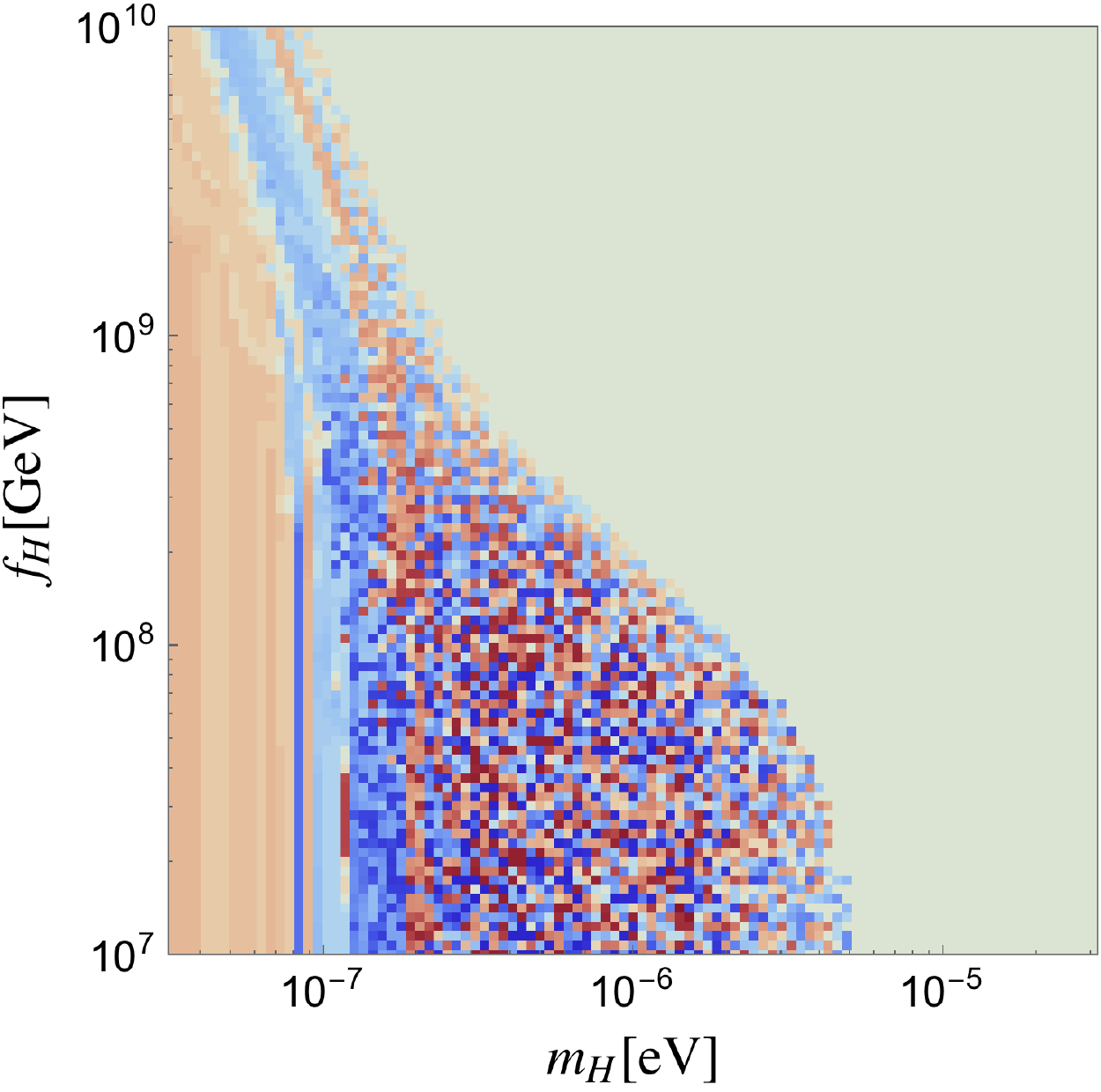}
  \label{fig6}
  }&
  \subfigure{
\includegraphics[height=6cm]{density4.pdf}
  \label{density2}
  }  
 \end{tabular}
  \caption{
  Same as Fig.~\ref{Fa12} but for $F_a = 10^{10}$\,GeV and $(\theta_i,n_a) = (1.5,5), \,(2.5,5)$ and $(1.5,15)$
  for which $m_a(T_{\rm osc}) \simeq  4 \times 10^{-8},\,1 \times 10^{-7},\,4 \times 10^{-8}$\,eV, respectively. 
   }
   \label{Fa10}
   \end{figure}

Similarly, the case with $F_a = 10^{10}$\,GeV is shown in Fig.~\ref{Fa10}. 
The condition, $f_H < F_a/n_a$, reads $f_H \lesssim 2 \times 10^{9}$\,GeV in Figs.~\ref{fig4} and \ref{fig5},
and $f_H \lesssim 7 \times 10^{8}$\,GeV in Fig.~\ref{fig6}, respectively.
As expected from the conditions (\ref{cond1})
and (\ref{condition1}) (and (\ref{hratio})), the multicolored region is shifted to larger $m_H$ and  smaller $F_a$.

%
%

\section{Cosmological implications} 
\label{implication}
Once the axion roulette takes place, domain walls are likely produced as $\Theta_f$ is extremely sensitive to
the initial misalignment angle $\theta_i$. We have confirmed by numerical calculations that $\Theta_f$ takes different values
even if $\theta_i$ has a small fluctuation of order $\delta \theta_i \sim 10^{-5}$.
One solution to the cosmological domain wall  problem is to invoke  late-time inflation 
 to dilute the abundance of domain walls. In our case, however, this is unlikely because the domain walls are formed at the
 QCD phase transition, and it is highly non-trivial to realize sufficiently long inflation and successful baryogenesis
 at such low temperatures.  Another is to make domain walls unstable and quickly decay
by introducing energy bias between different vacua.\footnote{
It is also possible that $\Lambda_H$ is time-dependent and it vanishes in the present Universe.
Then the energy density of domain walls becomes negligible, avoiding the cosmological domain wall problem.
} Note however that one cannot introduce any energy
bias between the vacua that are identical to each other (i.e. $\Theta_f = 0$ and $\Theta_f = 2 \pi n_H m$ with $m \in {\bf Z}$).\footnote{
 Here we assume that the QCD axion is fixed at the same minimum with $a_H$  differing from vacuum to vacuum.
} So, 
if both vacua with $\Theta_f = 0$ and $\Theta_f = 2 \pi n_H$ are populated in space, the domain walls connecting them
 are stable and cannot be removed even if one introduces energy bias between different vacua.\footnote{
  One may avoid this problem by considering a monodromy-type energy bias term.
 } This argument led us to conclude that the parameter region where the axion roulette occurs and $\Theta_f$ takes large positive or negative values is plagued with
cosmological domain wall problem, unless the spatial variation of $\Theta_f$ is much smaller than
$2\pi  n_H$. This requires either a large value of $n_H$ or negligible fluctuations of the initial misalignment angle $\delta \theta_i$.

In the following, let us consider a case where domain walls are formed, but the spatial variation of $\Theta_f$ is much 
smaller than $2 \pi n_H$. In this case, one may avoid the domain wall problem by introducing an energy bias between
different vacua. This corresponds to e.g. the left edge (the lower end of $m_H$) of the multicolored regions in Figs.~\ref{Fa12} and \ref{Fa10},
where the axion roulette takes place but the dependence on $\theta_i$ is relatively mild.
(See also Fig.~\ref{chaotic}.)
  
As a specific example, the bias term may be written as
 \beq
 V_{\rm bias}=\L'^4\left[1-\cos\left(N_H\frac{a_H}{F_H}+N_a\frac{a}{F_a}+\d\right)\right],
\eeq
 where $N_H$ and $N_a$ are integers, and $\delta$ is a CP phase. In the presence of the bias term,
 the minimum of the QCD axion is generally deviated from the CP conserving minimum. Depending on the size of  $\L'^4$ and 
 $\delta$,  the strong CP phase may exceed  the neutron electric dipole moment (EDM) constraint~\cite{Baker:2006},
 \beq
\bar\h \equiv \frac{\la a \ra}{F_a} <0.7\times10^{-11},\label{NEDM}
\eeq
which would spoil the PQ solution to the strong CP problem. On the other hand, if the magnitude of the bias term $( \L'^4)$
is too small, the domain walls become so long-lived that they may overclose the Universe or overproduce  axions
by their annihilation.
Therefore it is non-trivial if one can get rid of domain walls by 
energy bias without introducing a too large contribution to the strong CP phase or producing too many axions. 
Indeed, in the case of the QCD axion domain walls,  it is known that a mild tuning of the CP phase of the energy bias term 
is required~\cite{Kawasaki:2014sqa}.

To be concrete, let us  focus on the case of $N_H = 1$ and $N_a = 0$.
 Other choice of $N_H$ and $N_a$ does not alter our results significantly.
Assuming $V_{\rm bias}$ is a small perturbation to the original axion potential, i.e., $\Lambda'^4 \ll \Lambda_{H}^4 < m_a^2 F_a^2$, 
one can expand the total potential $V_{\rm QCD}+V_H+V_{\rm bias}$ around  $a=a_H=0$. Then we obtain 
\beq
\bar\theta \simeq  \frac{n_a \Lambda'^4}{n_H m_a^2 F_a^2} \sin \delta.
\label{thetab}
\eeq
Thus, the strong CP phase is induced by the bias term.
Requiring that $\bar\h$ should not exceed the neutron EDM constraint (\ref{NEDM}),
we obtain an upper bound on $\L'$ for given $\delta$. For $\delta = {\cal O}(1)$,
$\Lambda'$ must be smaller than the QCD scale by more than a few orders of magnitude.

The QCD axion and the ALP contribute to dark matter. 
In the absence of the mixing, the abundance of the QCD axion from the misalignment mechanism 
is  given by~\cite{Turner:1986}
\beq
\Omega_ah^2=0.18\,\h_i^2\left(\frac{F_a}{10^{12}~{\rm GeV}}\right)^{1.19}\left(\frac{\L_{\rm QCD}}{400~ {\rm MeV}}\right),
\label{axiondensity}
\eeq
where we have neglected the anharmonic effect and $h \simeq 0.7$ is the dimensionless Hubble
parameter. In the presence of the mixing with an ALP, a part of the initial oscillation energy
turns into the kinetic energy of the ALP, if the axion roulette is effective.  According to our numerical calculation, the QCD axion abundance
decreases by several tens of percent when  the axion roulette takes place. 

Next, we consider the ALP production. The ALP is mainly produced by the annihilation of domain walls.
Assuming the scaling behavior, the domain wall energy density is given by
\beq
\r_{\rm DW}\sim\s H,
\eeq 
where $\s\simeq 8m_H f_H^2$ is the tension of the domain wall, and $H$ is the Hubble parameter. The domain walls annihilate
when their energy density becomes comparable to the bias energy density, $
\rho_{\rm DW} \sim \Lambda'^4$. The produced ALPs are only marginally relativistic, and they become soon non-relativistic
due to the cosmological redshift. The ALP abundance is therefore
\beq
\Omega_{\rm ALP} h^2 \;\simeq\; 0.4\,\lrfp{m_H}{10^{-7}\,{\rm eV}}{\frac{3}{2}}
\lrfp{f_H}{10^{10}\,{\rm GeV}}{3} \lrfp{\Lambda'}{1 {\rm \, keV}}{-2},
\eeq
where we have set $g_*(T) = 10.75$. In order not to exceed the observed dark matter abundance $\Omega_c h^2 \simeq 0.12$~\cite{Ade:2015xua},
the size of the energy bias is bounded below:
\beq
\Lambda' \;\gtrsim\; 2 {\rm\,keV}\, \lrfp{m_H}{10^{-7}\,{\rm eV}}{\frac{3}{4}}
\lrfp{f_H}{10^{10}\,{\rm GeV}}{\frac{3}{2}}. \label{DM}
\eeq

There is another constraint coming from the isocurvature perturbations. In general, domain walls are formed when the corresponding scalar 
field has large spatial fluctuations. Once the domain wall distribution reaches the scaling law, isocurvature perturbations of domain walls
are suppressed at superhorizon scales. However, those ALPs produced during or soon after the domain wall formation are considered
to have sizable fluctuations at superhorizon scales, which may contribute to the isocurvature perturbations.  
The energy density of such ALPs at the domain wall formation is estimated to be
\beq
\delta \r_{\rm ALP, osc} \sim m_H^2f_H^2.
\eeq
Then the CDM isocurvature perturbation is 
\beq
\delta_{\rm iso} =\frac{\d\rho_{\rm ALP}}{\r_c}\sim\frac{\Omega_{\rm ALP}}{\Omega_c} \frac{m_H^2f_H^2}{\s H_{\rm ann}}\left(\frac{a_{\rm osc}}{a_{\rm ann}}\right)^3,
\eeq
where $\rho_c$ is the CDM energy density. Assuming that the Universe is radiation dominated at the domain wall formation, the CDM isocurvature is expressed as
\beq
\delta_{\rm iso} \sim 2\times10^{-4}\left(\frac{m_H}{10^{-7}~{\rm eV}}\right)^2\left(\frac{f_H}{10^{10}~{\rm GeV}}\right)^2\left(\frac{H_{\rm osc}}{10^{-9}~{\rm eV}}\right)^{\frac{3}{2}}.
\eeq
The Planck 2015 constraint on the (uncorrelated) isocurvature perturbations gives $\delta_{\rm iso} \lesssim 9.3 \times 10^{-6}$~\cite{Ade:2015lrj} and we obtain
\beq
\left(\frac{m_H}{10^{-7}\,{\rm eV}}\right)\left(\frac{f_H}{10^{10}\,{\rm GeV}}\right)\lesssim 9 \times10^{-2},\label{iso1}
\eeq
where we set $F_a = 10^{10}$\,GeV to evaluate $H_{\rm osc}$. 
For $F_a=10^{12}$ GeV, it reads
\beq
\left(\frac{m_H}{10^{-7}\,{\rm eV}}\right)\left(\frac{f_H}{10^{10}\,{\rm GeV}}\right)\lesssim 3 \times10^{-1}.\label{iso2}
\eeq

In the above, we have focused only on the linear perturbation for the isocurvature perturbation. However, since the spatial fluctuation 
of ALP becomes O(1) after the axion roulette, the higher order terms can also be significant and the isocurvature perturbation becomes highly non-Gaussian.
In this case, the non-Gaussianity is estimated as 
$\alpha^2 f_{\rm NL}^{\rm (iso)} \sim 
160 (\delta_{\rm iso}/9.3 \times 10^{-6})^3$ \cite{Kawasaki:2008sn,Langlois:2008vk},
which should be compared with the current 2-$\sigma$ constraint $|\alpha^2 f_{\rm NL}^{\rm (iso)}| < 140$ \cite{Hikage:2012be}.\footnote{
As pointed out in Ref.~\cite{Hikage:2012be}, the constraint should be regarded as a rough estimate when the quantum fluctuations dominate
over the classical field deviation from the potential minimum.} Therefore, the  non-Gaussianity constraint is comparable to that from the isocurvature
perturbations power spectrum.

\begin{figure}
\centering
\includegraphics[width=10cm]{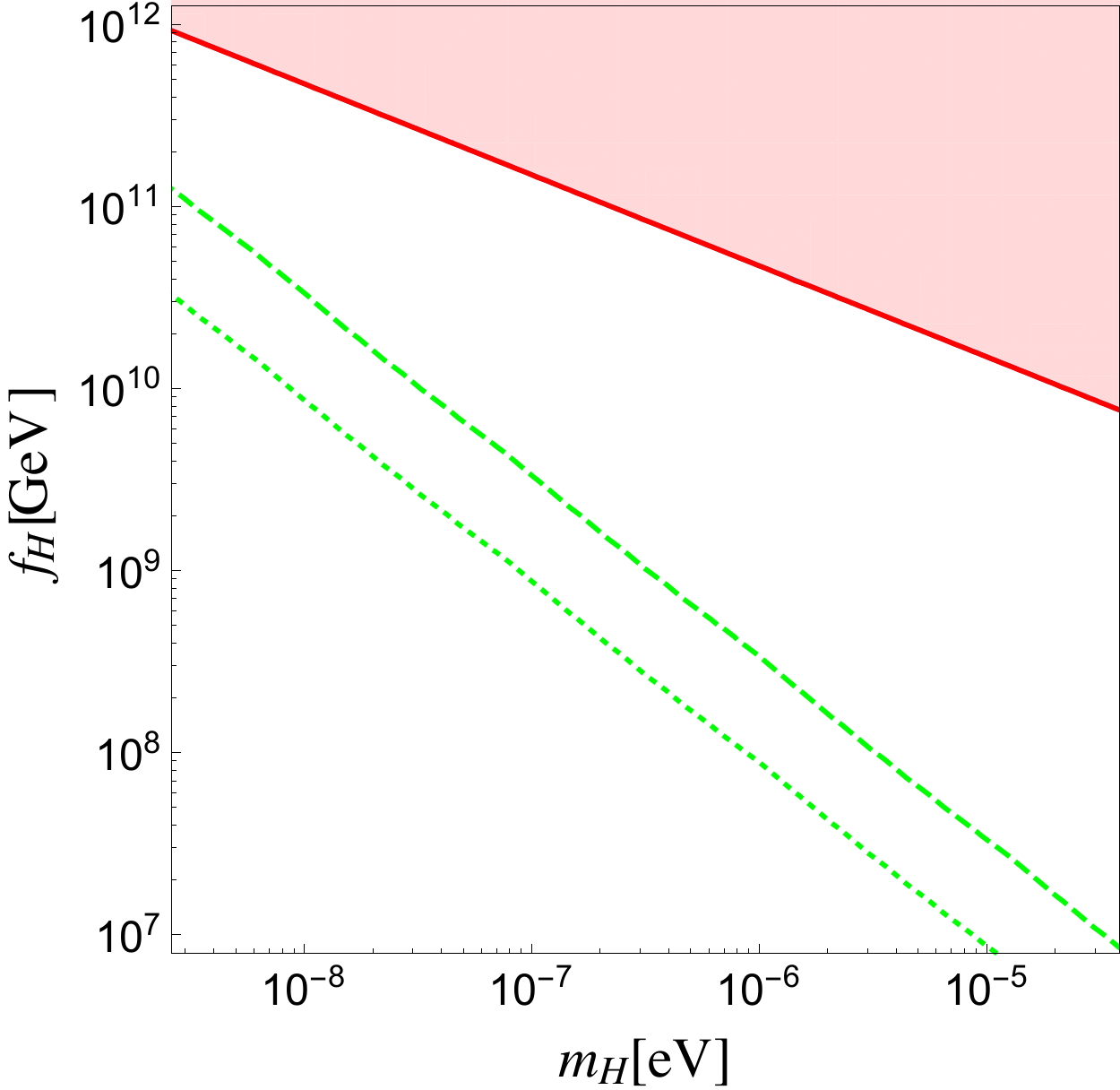}
\caption{
Upper bounds on  $m_H$ and $f_H$ from the DM abundance  and the neutron EDM constraint. 
 Here we set the phase of the bias term $\d=1$, and the domain wall numbers $n_H=2,~n_a=5$.
The shaded region above the 
solid (red) line is excluded because no $\L'$ can satisfy both (\ref{NEDM}) and (\ref{DM}) simultaneously. The dashed (dotted) green line denotes isocurvature bound for $F_a=10^{12} (10^{10})$ GeV. 
 }
\label{constraint}
\end{figure}

In Fig.~\ref{constraint} we show the upper bounds on $m_H$ and $F_H$ from the neutron EDM constraint (\ref{NEDM}) with (\ref{thetab}),
and isocurvature perturbations (\ref{iso1}). Compared to Figs.~\ref{Fa12} and \ref{Fa10}, one can see that there are allowed regions where the axion roulette takes place and the upper bounds are satisfied. Such regions are cosmologically allowed even if domain walls are formed
through the axion roulette, because the domain walls are unstable and decay quickly without spoiling the PQ mechanism.

\section{Conclusions}
\label{conc}
In this paper we have studied in detail the level crossing phenomenon between the QCD axion and an ALP,
focusing on the recently found {\it axion roulette}, in which the ALP runs along the valley of the potential,
passing through many crests and troughs before it gets trapped at one of the potential minima. Interestingly,
the axion dynamics shows rather chaotic behavior, and it is likely that domain walls (without boundaries) are
formed. We have determined the parameter space where the axion roulette  takes place and it is
represented by the multicolored regions in  Figs.~\ref{Fa12} and \ref{Fa10}.
As the domain  walls are cosmological stable, such parameter region does not lead to viable cosmology. In a certain case,
the domain walls can be made unstable by introducing an energy bias between different vacua, and we have
estimated the abundance of the ALPs dark matter produced by the domain wall annihilation. 
In contrast to the QCD axion domain walls, there is a parameter space where no fine-tuning of the CP phase of the 
bias term is necessary to make domain walls decay rapidly.

\section*{Acknowledgment}
This work is supported by MEXT Grant-in-Aid for Scientific research 
on Innovative Areas (No.15H05889 (F.T.) and No. 23104008 (N.K. and F.T.)), 
Scientific Research (A) No. 26247042 and (B) No. 26287039 (F.T.), 
and Young Scientists (B) (No. 24740135 (F.T.)),
and World Premier International Research Center Initiative (WPI Initiative), MEXT, Japan (F.T.).
N.K. acknowledges the Max-Planck-Gesellschaft, the Korea Ministry of Education, Science and Technology, 
Gyeongsangbuk-Do and Pohang City for the support of the Independent Junior Research Group at the 
Asia Pacific Center for Theoretical Physics.




\end{document}